\documentclass[a4paper,11pt]{article}
\usepackage{jheppub} % for details on the use of the package, please
\usepackage{bm}
\usepackage{soul}
\usepackage{xcolor}

\usepackage[T1]{fontenc} % if 

\newcommand\mb {\bar{\mu}}

\title{\boldmath Computing $c$- and $a$-functions from entanglement}

\author[a]{K. Boutivas,}
\author[a]{D. Katsinis,}
\author[a,b]{G. Pastras,}
\author[a]{and N. Tetradis}

\affiliation[a]{Department of Physics, University of Athens, Zographou 157 84, Greece}
\affiliation[b]{Laboratory for Manufacturing Systems and Automation, Department of Mechanical Engineering and Aeronautics, University of Patras, Patra 26110, Greece}

\emailAdd{kboutivas@phys.uoa.gr}
\emailAdd{dkatsinis@phys.uoa.gr}
\emailAdd{pastras@lms.mech.upatras.gr}
\emailAdd{ntetrad@phys.uoa.gr}

\abstract{
We confirm the direct connection between entanglement entropy and the notion of irreversibility in the renormalization-group flow in the context of a simple theory for which a calculation from first principles is feasible. The change of the entanglement  entropy for a spherical entangling surface as its radius grows from zero to infinity corresponds to the flow from the UV to the IR. Through analytical and numerical means, we compute the entanglement entropy for a free massive scalar theory, making use of the method of correlation functions. We deduce  a $c$-function in $1+1$ dimensions and an \mbox{$a$-function} in $3+1$ dimensions. Both functions are monotonic and vary continuously between one and zero, as expected for this simple theory. 
}

\begin{document} 
\maketitle
\flushbottom

\section{Introduction}\label{sec:intro}
The irreversibility of the renormalization-group (RG) flow \cite{Wilson:1973jj} is formalized in $1+1$ dimensions through the celebrated $c$-theorem \cite{Zamolodchikov:1986gt}, which demonstrates that there exists a monotonically decreasing $c$-function that interpolates between the central charge of a conformal field theory (CFT) at an ultraviolet (UV) fixed point and the corresponding central charge at an infrared (IR) fixed point at the end of the flow. We are interested in generalizations of the $c$-function in higher dimensions. In particular, we focus on the $a$-function in $3+1$ dimensions \cite{Cardy:1988cwa,Jack:1990eb,Komargodski:2011vj,Hartman:2023qdn}, which interpolates between the coefficients of the $A$-type conformal anomaly, associated with the Euler density, for the CFTs at the two ends of the RG flow. We present the calculation of such an $a$-function in a very simple setup, which allows a direct construction from first principles.

We exploit the connection between the $a$-function and the entanglement entropy of a subsystem consisting of the part of a quantum field theory enclosed by a spherical entangling surface of radius $R$. A key aspect of this connection is that the entanglement entropy of this theory at its vacuum state as a function of $R$ can be used in order to define appropriate $c$- and $a$-functions in flat \cite{Casini:2004bw,Casini:2005zv,Casini:2012ei,Casini:2017vbe} or curved \cite{Abate:2024nyh} spaces.\footnote{An $F$-function in $2+1$ dimensions can also be defined. We consider only spaces with an even number of dimensions in this work, for which the values of the function at the endpoints of the flow have a clear universal interpretation \cite{Casini:2012ei}.} The flow from the UV to the IR corresponds to the evolution of the entanglement entropy when $R$ increases from zero to infinity. 

We consider the simplest possible system, that of a free massive scalar field in $1+1$ and $3+1$ dimensions. In both cases the UV fixed point corresponds to the massless theory, while the field decouples in the IR, so that no degrees of freedom survive in this limit. In $1+1$ dimensions, the entanglement entropy at the UV fixed point displays the well known logarithmic dependence on the size of the subsystem, with a coefficient that is proportional to the central charge of the corresponding CFT \cite{Holzhey:1994we,Korepin:2004zz,Calabrese:2004eu}. For the simple theory that we consider, we expect that the $c$-function interpolates monotonically between central charges equal to 1 and 0 at the two ends of the flow. Such a function was computed in \cite{Casini:2009sr}. In $3+1$ dimensions the $a$-function should reproduce the coefficient of the $A$-type conformal anomaly in the UV \cite{Solodukhin:2008dh,Lohmayer:2009sq,Casini:2009sr,Casini:2010kt}, while it should vanish in the IR. 

A fully analytical calculation of entanglement entropy in $3+1$ dimensions is hindered by the difficulty in implementing techniques such as the replica method. However, a calculation from first principles is still possible through the computation of the eigenvalues of the reduced density matrix \cite{Sorkin:1984kjy,Bombelli:1986rw,Srednicki:1993im}, or the covariance matrix built from correlation functions of the subsystem \cite{Peschel:2002yqj,Casini:2009sr,Sorkin:2012sn}. The determination of the relevant part of the entropy requires the subtraction of the UV divergences of the fixed-point theory, as well as the ones resulting from relevant perturbations \cite{Casini:2012ei,Casini:2017vbe,Abate:2024nyh}. Our recent work on the entanglement entropy in various gravitational backgrounds \cite{Boutivas:2024lts,Boutivas:2024sat,Boutivas:2025ksp,Boutivas:2025rdf} has led to a sufficient understanding of the UV divergent terms, as well as possible IR effects. In the simple model we consider here, it is then possible to  isolate the finite part.

In this work we employ the finite part of the entanglement entropy for the computation of $c$- and $a$-functions. The relevant expressions are \cite{Casini:2012ei,Casini:2017vbe,Abate:2024nyh}
\begin{align}
	\Delta  C(R)&= R\, \Delta S'(R) ~~~~~~~~~~~~~~~~~~~~~~~~{\rm in\,\, 1+1 \, dimensions} \label{eq:dcr} \\
	\Delta  A(R)&= R^2\, \Delta S''(R)-R\, \Delta S'(R) ~~~~~~{\rm in \,\, 3+1 \, dimensions}	. \label{eq:dar}
\end{align}
The quantity $\Delta S(R)$ is obtained after the subtraction of the entanglement entropy at the UV fixed point, as well as possible UV divergences in the deformed theory resulting from relevant perturbations. In our case, these are UV-divergent terms that are proportional to the field mass. As a result, the functions $\Delta  C(R)$ and $\Delta  A(R)$ vanish at $R=0$, where the entropy is dominated by the contributions from the UV fixed point. 

The structure of the paper is as follows: In section~\ref{sec:method} we summarize the main elements of our methodology. In section~\ref{sec:cfunction} we compute numerically the $c$-function. We also present an analytical result for its form near $R=0$, with the details of the calculation given in the appendix. In section~\ref{sec:afunction} we present our results for the $a$-function. Finally, section~\ref{sec:concl} contains our conclusions.

\section{Methodology}\label{sec:method}

The formalism for the calculation of the entanglement entropy has been presented in several publications. We list here only the basic elements, with more details given in \cite{Srednicki:1993im,Lohmayer:2009sq,Huerta:2022tpq,Boutivas:2024lts}. We consider a free massive real scalar field in $(3+1)$-dimensional flat space. Since we are interested in spherical entangling surfaces, we expand the field in real spherical harmonic moments
\begin{equation}
\phi_{\ell m} \left( r \right) = r \int {d\Omega \, Y_{\ell m} \left( {\theta ,\varphi} \right) \phi \left( \mathbf{x} \right)},
\end{equation}
and similarly for the conjugate momentum. In this equation $ Y_{\ell m} \left( {\theta ,\varphi} \right)$ stands for the real spherical harmonics. The radial coordinate is discretized through the introduction of a lattice of concentric spherical shells with radii $r_j = j \epsilon$, where $j = 1 , 2 , \ldots , N$. The UV-cutoff energy scale is $1 / \epsilon$, while the size of the lattice $L = N \epsilon$ sets an IR-cutoff energy scale $1 / L$. We define the canonically commuting, discretized degrees of freedom as $\phi_{\ell m} \left( {j \epsilon} \right) \to \phi_{\ell m,j}$ and $\pi_{\ell m} \left( {j \epsilon} \right) \to {\pi_{\ell m,j}}/{\epsilon}$. The Hamiltonian reads
\begin{equation}
	H = \frac{1}{2 \epsilon} \sum\limits_{\ell,m} \sum\limits_{j = 1}^N \Bigg[ \pi_{\ell m,j}^2 
	 + \left(\phi_{\ell m,j + 1}- \phi_{\ell m,j} \right)^2 
	+  \frac{\ell\left( {\ell + 1} \right)}{j^2}   \phi_{\ell m,j}^2
	+ \mb^2 \, \phi_{\ell m,j}^2  \Bigg]\equiv \sum\limits_{\ell,m} H_{\ell m} ,
	\label{eq:Hamiltonian_discretized_flat} 
\end{equation}
with $\mb=\mu\epsilon$. We express all dimensionful quantities in units of $\epsilon$, which is equivalent to setting $\epsilon=1$ in the numerical calculation. It is important to note that this form of the discretized Hamiltonian implements Dirichlet boundary conditions, with vanishing spherical moments $\phi_{\ell m} \left( r \right)$ at the origin and at a radius  $r=L$. These conditions stem from the smoothness of the original field at the origin and its boundary conditions at infinity. As a result, possible zero modes are eliminated, even though they may be present for a different choice of boundary conditions. 

Equation \eqref{eq:Hamiltonian_discretized_flat} shows that moments with different angular momenta do not interact. Moreover the dynamics of each sector of the Hamiltonian $H_{\ell m}$ depends on $\ell$ solely. As a result, taking into account the degeneracy, the entanglement entropy results from the addition of contributions from all $\ell$-sectors, and reads
\begin{equation}\label{eq:SEE_ell_sum}
	S\left( n , N, \mb \right) = \sum_{\ell=0}^{\infty} \left( 2 \ell + 1 \right) S_\ell \left( n , N, \mb \right) ,
\end{equation}
where $S_{\ell}$ is the entropy for a single $\ell$-sector, which has been split into the subsystem $A$, containing $\phi_{\ell m,j}$ with $j \leq n$, and its complementary subsystem $A^C$. This separation corresponds to a spherical entangling surface of radius $R = \left( n + \frac{1}{2} \right) \epsilon\equiv n_R \, \epsilon.$

When the overall system lies at its ground state, the state of each $\ell$-sector reads
\begin{equation} \label{eq:wavefunction}
	\Psi({{\bm{\phi}}_{\ell\vec{m}}})=\left(\det\frac{\Omega}{\pi}\right)^{1/4} \exp\left(-\frac{1}{2} {\bm{\phi}}_{\ell\vec{m}}^T \Omega\, {\bm{\phi}}_{\ell\vec{m}}\right) ,
\end{equation}
where $\Omega$ is the positive square root of the coupling matrix $K$ of the corresponding $\ell$-sector Hamiltonian, and  ${\bm{\phi}}_{\ell\vec{m}}$ is a column vector containing $\phi_{\ell\vec{m},j}$. The matrix $K$ can be directly read from equation \eqref{eq:Hamiltonian_discretized_flat}, namely
\begin{equation}\label{eq:Coupling_Matrix}
	K_{ij} = \left( 2 + \mu^2\epsilon^2 + \frac{\ell \left( \ell + 1 \right) }{i^2}\right) \delta_{i , j}   - \delta_{i + 1 , j} - \delta_{i , j + 1}.
\end{equation}
The $\ell=0$ sector corresponds to the discretized Hamiltonian of the $(1+1)$-dimensional field theory. The system can be split into blocks by writing
\begin{equation}\label{eq:blocks}
	\Omega = \left( \begin{array}{cc} \Omega_{A} & \Omega_{B} \\ \Omega_B^T & \Omega_C \end{array} \right) , \quad {\bm{\phi}} = \left( \begin{array}{c} {\bm{\phi}}_A \\ {\bm{\phi}}_C \end{array} \right) ,
\end{equation}
where the vector ${\bm{\phi}}_A$ consists of the field values at the lattice points of subsystem $A$, and ${\bm{\phi}}_C$ of those of its complement $A^c$.

In the original approach of \cite{Srednicki:1993im} the entanglement entropy was computed through the reduced density matrix, which reads
\begin{equation}
	\rho_A \left( {\bm{\phi}}_A ; {\bm{\phi}}_A^\prime \right) \sim \exp \bigg[ - \frac{1}{2} \left( {\bm{\phi}}_A^T \gamma {\bm{\phi}}_A  + {\bm{\phi}}_A^{\prime T} \gamma {\bm{\phi}}^\prime_A \right)  + {\bm{\phi}}_A^{\prime T} \beta {\bm{\phi}}_A \bigg] ,\label{eq:reduced_density_matrix}
\end{equation}
with
\begin{equation}
	\gamma = \Omega_A - \frac{1}{2} \Omega_B^T \Omega_C^{-1} \Omega_B , \quad 
	\beta = \frac{1}{2} \Omega_B^T \Omega_C^{-1} \Omega_B .
	\label{eq:gamma_beta}
\end{equation}
The contribution of a single $\ell$-sector to the entanglement entropy is
\begin{equation}
	S_{\ell} = - \sum_{i = 1}^n \left( \ln \left( 1 - \xi_i \right) + \frac{\xi_i}{1 - \xi_i} \ln \xi_i \right) ,
	\label{eq:SEE}
\end{equation}
where $\xi_i = \frac{\beta_i}{1 + \sqrt{1 - \beta_i^2}}$, with $\beta_i$ the eigenvalues of the matrix $\tilde{\beta} = \gamma^{-1} \beta$. 

The above approach is equivalent to the covariance matrix method \cite{Peschel:2002yqj,Casini:2009sr,Sorkin:2012sn,Katsinis:2024gef}. It can be shown that the entanglement entropy is given as
\begin{equation}\label{eq:SEE_of_M_cal}
	S_\ell = \sum_{i=0}^{n} \left( \frac{\sqrt{\lambda_i} + 1}{2} \ln \frac{\sqrt{\lambda_i} + 1}{2} - \frac{\sqrt{\lambda_i} - 1}{2} \ln \frac{\sqrt{\lambda_i} - 1}{2} \right),
\end{equation}
where $\lambda_i$ are the eigenvalues of the matrix
\begin{equation}\label{eq:mat_M_tilde}
\mathcal{M}=( \Omega^{-1} )_A  (\Omega)_A.
\end{equation}

In the following sections we present high-precision numerical calculations of the entanglement entropy for the $(1+1)$- and $(3+1)$-dimensional massive scalar field theory in flat space, using the method of correlation functions. A detailed description of the numerical process is given in \cite{Lohmayer:2009sq,Boutivas:2024lts}. The $(1+1)$-dimensional theory corresponds to the vanishing-angular-momentum sector of the $(3+1)$-dimensional theory. The analysis consists of the following steps:
\begin{enumerate}
\item We determine and subtract the finite-size effects for each angular-momentum sector with $\ell\leq 220$. To this purpose, we analyze the data as functions of $N$ for fixed values of $n$ and $\mb$.The appropriate form of the expansion is
\begin{equation}\label{eq:vac_3d_N_corrections}
		S_{\ell}(n,N,\mb)=S_{\ell,\infty}(n,\mb)+\sum_{k=0}^{k_{\textrm{max}}}\frac{S_{\ell}^{(k)}(n,\mb)}{N^{(2\ell+2+k)}}.
\end{equation}
For $\ell\geq 221$ the finite-size effects are considered negligible. Even though we take into account angular-momentum sectors up to $\ell \sim 10^6$, the finite-size analysis is performed only for $\ell \leq 220$. To obtain results for the continuous theory in the infinite-size limit, we use the coefficient $S_{\ell,\infty}(n,\mb)$.
\item In order to account for the contributions of all angular-momentum sectors, we calculate the truncated sum 
\begin{equation}\label{eq:trunc_sum_vac}
		S_{\infty}(n,\mb;\ell_\textrm{max})=\sum_{\ell=0}^{\ell_\textrm{max}}(2\ell+1)S_{\ell,\infty}(n,\mb)
\end{equation}
for various values of $\ell_\textrm{max}$. This sum behaves as
\begin{equation} \label{eq:lmax_exp}
		S_{\infty}(n,\mb;\ell_\textrm{max})=S_{\infty}(n,\mb) + \sum_{i=1}^{i_\textrm{max}}\frac{1}{\ell_\textrm{max}^{2i}}(a_i(n,\mb)+b_i(n,\mb)\ln{\ell_{\textrm{max}}}).
\end{equation}
The quantity $S_{\infty}(n,\mb)$ corresponds to the limit $\ell_\textrm{max}\rightarrow\infty$.
\item Finally, we analyze the dependence of $S_{\infty}(n,\mb)$ on $n$ and $\mb$, or equivalently on $R = \left( n + \frac{1}{2} \right) \epsilon$ and $\mu=\mb/ \epsilon$. This quantity contains both UV-divergent and finite parts. For the case of a scalar field that we consider, it has the form (we omit the subscript $\infty$ in the following)
\begin{align}
		S(R,\mu)&= \frac{1}{6} \ln \frac{R}{\epsilon} + S_{\rm fin}(\mu R)~~~~~~~~~~~~~~~~~~~~~~~~~~~~~~~~~{\rm in\,\, 1+1 \, dimensions}, \label{eq:totentr11}\\
	S(R,\mu)&=c\frac{R^2}{\epsilon^2}-\frac{1}{90}\ln \frac{R}{\epsilon} +\frac{1}{6}\mu^2R^2\ln \mu\epsilon + S_{\rm fin}(\mu R) ~~~{\rm in \,\, 3+1 \, dimensions}	,
			\label{eq:totentr31}
\end{align}
where we dropped terms that vanish in the continuum limit $\epsilon\rightarrow0$. The coefficient of the logarithm in the first expression corresponds to a central charge $c=1$ at the UV fixed point in $1+1$ dimensions. The coefficient of the area term in the result for the $(3+1)$-dimensional theory depends on the choice of UV regulator. The coefficient of the second term in the same expression corresponds to the contribution of a real scalar field to the $A$-type conformal anomaly \cite{Birrell:1982ix,Casini:2009sr}. The third term encodes the UV divergence induced by the mass perturbation. Its coefficient was  computed in \cite{Hertzberg:2010uv} and is in agreement with numerical results for the same term in curved backgrounds \cite{Boutivas:2025ksp,Boutivas:2025rdf}. The finite parts depend only on the dimensionless product $\mu R$. 
\end{enumerate}	
It is important to note that, if the expressions \eqref{eq:dcr}, \eqref{eq:dar} are applied to the divergent parts, they isolate only the coefficients of the terms $\sim \ln R/\epsilon$, which are proportional to the central charge of the $(1+1)$-dimensional theory and the coefficient of the $A$-type conformal anomaly in the $(3+1)$-dimensional theory. This observation indicates a straightforward procedure in order to define $c$- and $a$-functions directly from the full entanglement entropy in our case. One can use as an independent variable $x=\mu R$ and express the entanglement entropy as $S(x,\mu)$. The quantities
\begin{equation}\label{eq:cx}
	c(x)= 6 x \,  S'(x,\mu)=1+ 6 x \,  S_{\rm fin}^{\prime}(x),
\end{equation}
in $(1+1)$-dimensions, and
\begin{equation}\label{eq:ax}
	a(x)= -45 \left( x^2\,  S''(x,\mu)-x\,  S'(x,\mu) \right)=1 -45 \left( x^2\,  S_{\rm fin}^{\prime\prime}(x)-x\,  S_{\rm fin}^{\prime}(x) \right),	
\end{equation}
in $(3+1)$-dimensions, define a $c$- and an $a$-function that are normalized so that they interpolate between 1 and 0 when $x$ increases from zero to infinity, independently of $\mu$. (The primes now indicate derivatives with respect to $x$.) More precisely, the dependence on $\mu$ disappears in the limit of vanishing cutoff $\epsilon$, which is equivalent to $\mb=\mu\epsilon \to 0$. Since $x$ is kept fixed in the process, we also have $R/\epsilon \to \infty$ in this limit. Even though we have also done the computation by first subtracting the UV divergences and using only the finite part, we have found that using the full entanglement entropy reduces the errors induced by the required fits in the intermediate steps of the subtraction procedure. The final results are the same for both procedures.  

It has been shown formally \cite{Casini:2004bw,Casini:2005zv,Casini:2012ei,Casini:2017vbe} that the expression \eqref{eq:ax} reproduces correctly the coefficients of the A-type conformal anomaly at the endpoints of the RG flow, which satisfy $a_{\textrm UV}>a_{\textrm UV}$. It must be noted, however, that the monotonicity along the flow is not guaranteed. Our calculation in section~\ref{sec:afunction} demonstrates this monotonicity in the context of the simple theory that we consider. Thus, it supports the conclusion that the expression \eqref{eq:ax} has the required properties of an $a$-function.

\section{The $c$-function}\label{sec:cfunction}
Before presenting the outcome of the numerical calculation, we quote some analytical results that will be very useful for checking its validity. Obtaining an analytical expression for the entire $c$-function is very difficult. However, it is possible to obtain approximate expressions in asymptotic regimes.

In order to study the behavior for $\mu R\to 0$, one can make use of the methodology for analytical calculations developed in \cite{Katsinis:2024gef}. Essentially, we take advantage of the fact that the entanglement entropy of the massless $(1+1)$-dimensional theory can be computed by considering the continuum limit of the expressions appearing in section \ref{sec:method}. In particular, using the eigenvalues of the kernel
\begin{equation}\label{eq:Kernel_M}
	\mathcal{M}\left(x,x^\prime\right)=\int_{R}^{L}dy\,\Omega^{-1}\left(x,y\right)\Omega \left(y,x^\prime\right),
\end{equation}
the entanglement entropy is calculated by applying equation \eqref{eq:SEE_of_M_cal}. The kernel $\mathcal{M}$ is the continuum limit of the matrix $\mathcal{M}$ defined in  equation \eqref{eq:mat_M_tilde}. In this equation $x$ and $x^\prime$ take values in $\left[0,R\right]$, while the overall system corresponds to the interval $[0,L]$.  For Dirichlet boundary conditions, using the spectral decomposition of the continuum limit of the coupling matrix \eqref{eq:Coupling_Matrix}, we have
\begin{align}
	\Omega\left(x,x^\prime\right)&=\frac{2}{L}\sum_{k=1}^{\infty}\left(\frac{k^2\pi^2}{L^2}+\mu^2\right)^{1/2}\sin\frac{k \pi x}{L}\sin\frac{k \pi x^\prime}{L},\label{eq:kernel_OM_sin}\\
	\Omega^{-1}\left(x,x^\prime\right)&=\frac{2}{L}\sum_{k=1}^{\infty}\left(\frac{k^2 \pi^2}{L^2}+ \mu^2\right)^{-1/2}\sin\frac{k \pi x}{L}\sin\frac{k \pi x^\prime}{L},\label{eq:kernel_OM_inv_sin}
\end{align}
where $x$ and $x^\prime$ take values in $\left[0,L\right]$. The eigenvalue problem of the kernel $\mathcal{M}$ can be solved exactly in the massless limit (even for finite size), and the solution can be used in order to perform perturbative calculations, see  \cite{Boutivas:2024sat,Boutivas:2025rdf}.

Here, we are interested in studying the RG flow triggered by the mass of the field in the infinite-size theory. To this purpose, starting from the massive, finite-size theory, we would like to first send the size of the system to infinity and then expand for small mass. Thus, we have to study the theory in the regime $\mu L\gg 1$. In order to take advantage of the result of \cite{Boutivas:2025rdf} and avoid redoing the entire analysis, in appendix~\ref{sec:analytic} we make use of the universality of the logarithmic terms. The structure of the leading correction to the entanglement entropy of the massless, infinite-size theory is the same, independently of whether we consider the small-mass expansion of the finite-size theory and then send the size to infinity ($m L \ll 1$), or we send first the system size to infinity and then expand for small mass ($m L \gg 1$). The results in the two limits can be mapped to each other through the appropriate identification of the IR regulators in the two cases. For $m L \gg 1$, in appendix~\ref{sec:analytic} we show that the entanglement entropy is given by
\begin{equation}
S(\mu R)=\frac{1}{6}\ln\frac{2R}{\epsilon}+\frac{1}{6}\mu^2R^2\left(\ln\left(\mu R\right)+\gamma -\frac{5}{6}\right)+\cdots,
\end{equation}
where $\gamma$ is the Euler–Mascheroni constant. As a result, the $c$-function reads
\begin{equation}\label{eq:correctionzero}
c(x)=1 + 2 x^2\left(\ln x + \gamma - \frac{1}{3}\right)+\cdots.
\end{equation}

\begin{figure*}[t]
	\centering
	\begin{picture}(91,53)
		\put(0.5,0){\includegraphics[angle=0,width=0.8\textwidth]{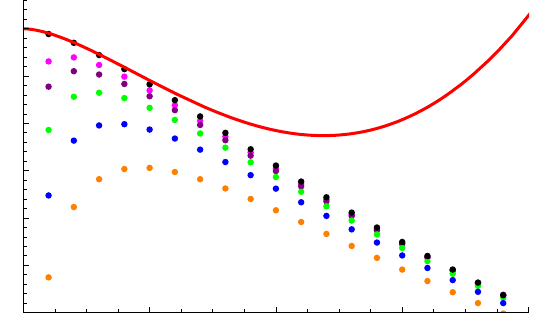}}
		\put(3.2,51.25){\Large $c$}
		\put(79,2.85){\Large $\mu R$}
		\put(02.5,0.35){\small $0.0$}
		\put(21.15,0.35){\small $0.2$}
		\put(39.80,0.35){\small $0.4$}
		\put(58.45,0.35){\small $0.6$}
		\put(77.1,0.35){\small $0.8$}
		\put(0,2.6){\small $0.4$}
		\put(0,9.6){\small $0.5$}
		\put(0,16.6){\small $0.6$}
		\put(0,23.6){\small $0.7$}
		\put(0,30.6){\small $0.8$}
		\put(0,37.6){\small $0.9$}
		\put(0,44.6){\small $1.0$}
		\put(75,09){\includegraphics[angle=0,width=0.05\textwidth]{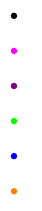}}
		\put(74.75,35){\includegraphics[angle=0,width=0.05\textwidth]{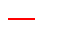}}
		\put(79,31.8){$\bar{\mu}= 0$}
		\put(79,27.6){$\bar{\mu}= 0.0025$}
		\put(79,23.4){$\bar{\mu}= 0.005$}
		\put(79,19.2){$\bar{\mu}= 0.01$}
		\put(79,15.0){$\bar{\mu}= 0.02$}
		\put(79,10.8){$\bar{\mu}= 0.04$}
		\put(79,36){Eq. \eqref{eq:correctionzero}}
		%\put(0,53){TESTV}
		%\put(91,28){TESTH}
	\end{picture}
\caption{The numerical results for the $c$-function for various values of $\mb=\mu\epsilon$. The black points  correspond to the extrapolation to the continuum limit $\epsilon\to 0$. The solid red line corresponds to the analytical expression \eqref{eq:correctionzero}, valid for small $\mu R$.}
\label{fig:c1}
\end{figure*}
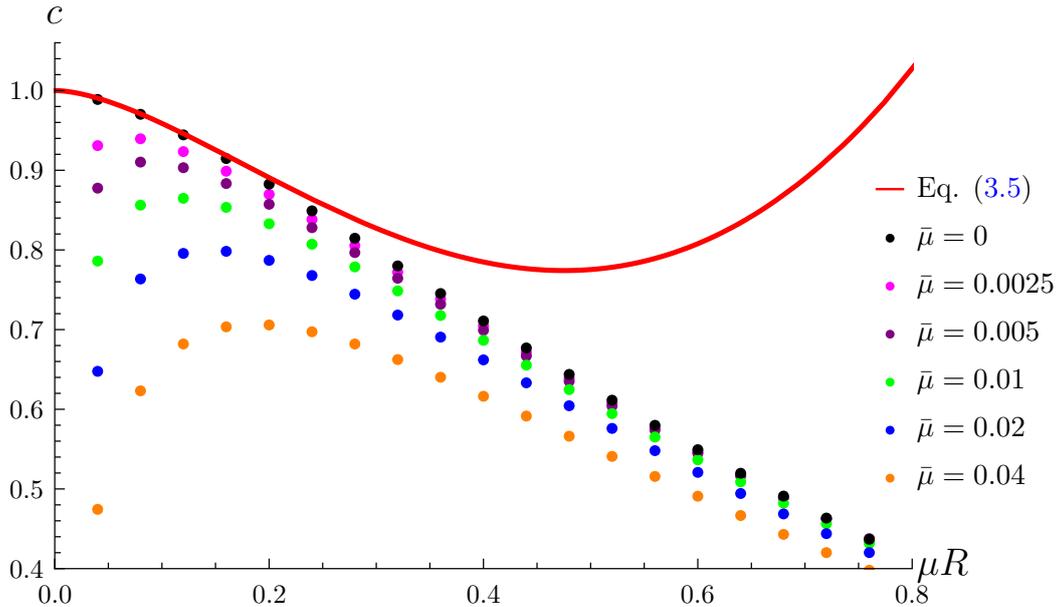

The above result indicates a smooth behavior near $\mu R=0$, with a vanishing derivative, in contrast to the result of \cite{Casini:2009sr}, which exhibits a cusp at this point. The cusp is related to an IR divergence in the entropy in the massless limit, which appears when a system of infinite size is studied with the replica method. On the other hand, we impose Dirichlet boundary conditions, with vanishing field at the endpoints of a finite system whose length is eventually taken to be infinite. This procedure eliminates possible zero modes from the system, as has been discussed in detail in \cite{Yazdi:2016cxn}, so that no strong IR effects emerge.

In the limit $\mu R\to \infty$ the entropy is expected to receive a correction 
\begin{equation}\label{eq:correctioninfinity}
S^{(\infty)}(\mu R)= -\frac{1}{6} \ln \mu R,
\end{equation}
which would combine with the first term in equation \eqref{eq:totentr11} in order to produce the expected $\sim \ln \mu \epsilon$ behavior of the non-critical system \cite{Calabrese:2004eu}. This behavior results in the vanishing of the $c$-function in this limit.

\begin{figure*}[t]
	\centering
	\begin{picture}(86,53)
		\put(1.5,0.15){\includegraphics[angle=0,width=0.8\textwidth]{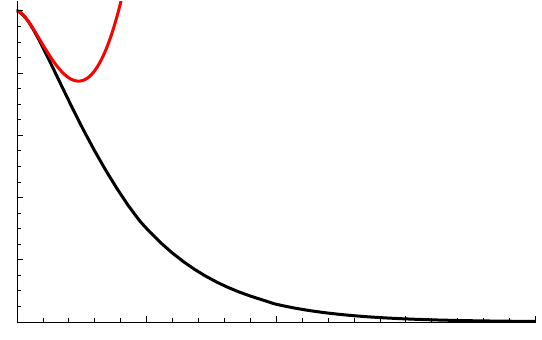}}
		\put(3.2,51.25){\Large $c$}
		\put(81.5,2.25){\Large $\mu R$}
		\put(03.500,0){\small $0$}
		\put(22.575,0){\small $1$}
		\put(41.675,0){\small $2$}
		\put(60.775,0){\small $3$}
		\put(79.875,0){\small $4$}
		\put(0,02.25){\small $0.0$}
		\put(0,11.35){\small $0.2$}
		\put(0,20.50){\small $0.4$}
		\put(0,29.60){\small $0.6$}
		\put(0,38.75){\small $0.8$}
		\put(0,48.00){\small $1.0$}
		\put(58,19){\includegraphics[angle=0,width=0.05\textwidth]{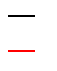}}
		\put(62,20){Eq. \eqref{eq:correctionzero}}
		\put(62,23.4){Numerical result}
		%\put(0,53){TESTV}
		%\put(86,3){TESTH}
	\end{picture}
\caption{The black line depicts the numerical result for the $c$-function in the continuum. The red line corresponds to the analytical expression \eqref{eq:correctionzero}, which is valid for small $\mu R$.}
\label{fig:c2}
\end{figure*}
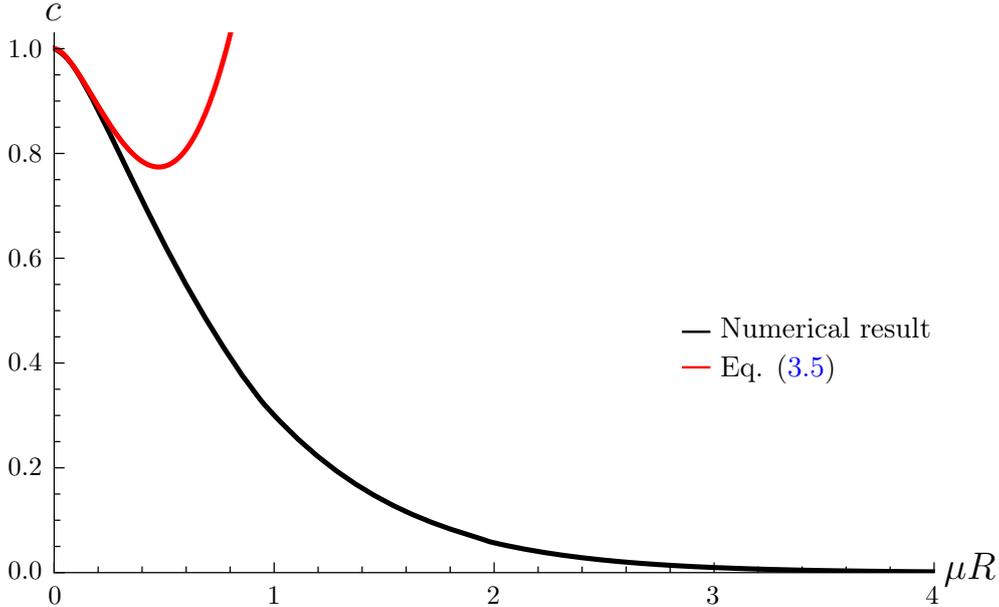

The numerical calculation of the $c$-function is performed as described in the section~\ref{sec:method}. The entanglement entropy is computed as a function of the radius $R$ for several values of the field mass $\mu$. It is then expressed as a function of $x=\mu R$ and the $c$-function is computed through equation \eqref{eq:cx}. As we discussed earlier, this relation is valid in the limit $\mb=\mu\epsilon \to 0$, in which any residual UV cutoff effects vanish. The practical limitation of a lattice of finite size implies that there is a maximal value for the ratio $R/\epsilon$ that we can consider. As a result, $\mu \epsilon$ cannot be taken arbitrarily small if the product $\mu R$ is to take values within a non-zero range. This issue becomes more severe for small values of $\mu R$. In fig. \ref{fig:c1} the colored points depict the outcome of the calculation of the $c$-function for a fixed maximal value of $R/\epsilon$ and various mass values, namely
\begin{equation}\label{eq:mass11}
	\mb=\mu\epsilon \in \left\{0.04, 0.02, 0.01, 0.005, 0.0025 \right\},
\end{equation}
starting from below. The convergence is apparent, but also incomplete, especially near $\mu R=0$. In order to approach the continuum limit, we perform a third-order polynomial fit for the values of the $c$-function in terms of $\mb$, for fixed $\mu R$. The extrapolation to $\mb=0$ returns the black points in the plot. The red continuous curve depicts the analytical result of equation \eqref{eq:correctionzero}. The agreement for the first points, in the region where this expression is expected to be a good approximation, is remarkable, and confirms that our calculation is robust.

The full $c$-function is depicted in fig. \ref{fig:c2}, along with the approximate expression \eqref{eq:correctionzero}. As expected, the function is monotonic. It starts from a value equal to 1 for $\mu R=0$ and approaches $0$ for large values of $\mu R$. Even though the vanishing of the $c$-function is expected on physical grounds, as the field decouples in this limit, it is a non-trivial feature from the point of view of the entanglement entropy. It means that the entropy satisfies equation 	\eqref{eq:correctioninfinity} exactly, a fact confirmed by the numerical analysis to a high precision.

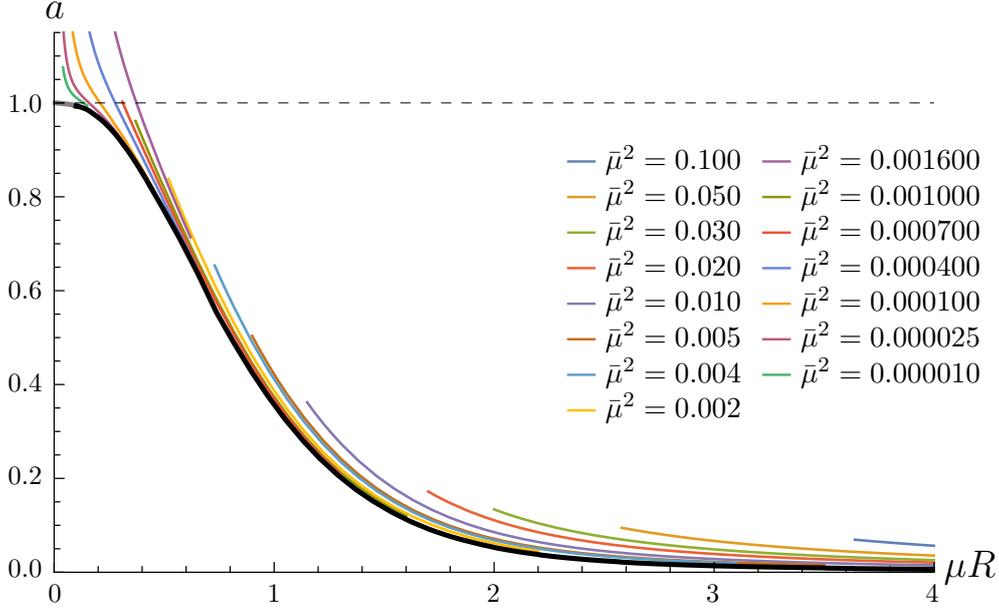
\begin{figure*}[t]
	\centering
	\begin{picture}(86,53)
		\put(1.5,0.15){\includegraphics[angle=0,width=0.8\textwidth]{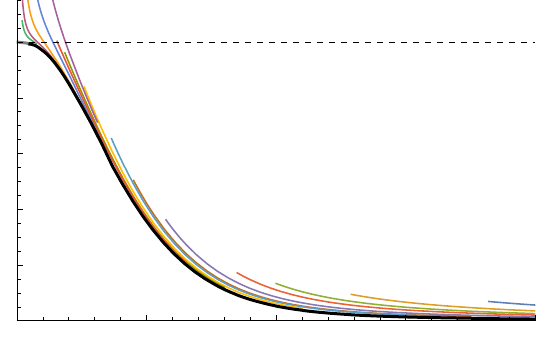}}
		\put(3.2,51.25){\Large $a$}
		\put(81.5,2.25){\Large $\mu R$}
		\put(03.500,0){\small $0$}
		\put(22.575,0){\small $1$}
		\put(41.675,0){\small $2$}
		\put(60.775,0){\small $3$}
		\put(79.875,0){\small $4$}
		\put(0,02.25){\small $0.0$}
		\put(0,10.35){\small $0.2$}
		\put(0,18.50){\small $0.4$}
		\put(0,26.60){\small $0.6$}
		\put(0,34.75){\small $0.8$}
		\put(0,43.00){\small $1.0$}
		\put(48,15.05){\includegraphics[angle=0,width=0.05\textwidth]{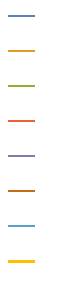}}
		\put(65,18.20){\includegraphics[angle=0,width=0.05\textwidth]{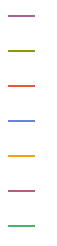}}
		\put(69,37.98){$\mb^2=0.001600$}
		\put(69,34.87){$\mb^2=0.001000$}
		\put(69,31.76){$\mb^2=0.000700$}
		\put(69,28.65){$\mb^2=0.000400$}
		\put(69,25.54){$\mb^2=0.000100$}
		\put(69,22.43){$\mb^2=0.000025$}
		\put(69,19.32){$\mb^2=0.000010$}
		\put(52,37.98){$\mb^2=0.100$}
		\put(52,34.87){$\mb^2=0.050$}
		\put(52,31.76){$\mb^2=0.030$}
		\put(52,28.65){$\mb^2=0.020$}
		\put(52,25.54){$\mb^2=0.010$}
		\put(52,22.43){$\mb^2=0.005$}
		\put(52,19.32){$\mb^2=0.004$}
		\put(52,16.21){$\mb^2=0.002$}
		%\put(0,53){TESTV}
		%\put(86,3){TESTH}
	\end{picture}
\caption{The colored lines depict the numerical results for the $a$-function for various values of $\mb=\mu\epsilon$. The black line corresponds to the extrapolation to the continuum limit $\epsilon\to 0$.}
\label{fig:a}
\end{figure*}

\section{The $a$-function}\label{sec:afunction}
 
We turn next to the $a$-function of the $(3+1)$-dimensional theory. Analytical results are not available for the entanglement entropy and the related $a$-function in this case. The only reasonable expectation, in analogy to the $(1+1)$-dimensional case, is that a term $\frac{1}{90}\ln \mu R$ must be present in the entropy for large $R$, so that the $a$-function vanishes for $R\to\infty$. However, a more dominant term $\sim \mu^2 R^2$, with an unknown coefficient, can also be present in the entropy, even though it does not affect the $a$-function defined in equation \eqref{eq:ax}. We do not perform a detailed analysis of the various terms in the entanglement entropy. We use instead the expression \eqref{eq:ax} for the direct computation of the $a$-function. 
 
The numerical calculation is performed as described in section \ref{sec:method}. It must account for the contributions of all $\ell$-sectors, which makes it much more demanding than the calculation of the $c$-function that involves only the $\ell=0$ sector. We depict the result in fig. \ref{fig:a}, in complete analogy to the presentation in fig. \ref{fig:c1}. The colored lines are obtained for the same range of $R/\epsilon$ and various mass values, namely
\begin{multline}\label{eq:mass31}
\mb^2  \in \left\{0.1, 0.05, 0.03, 0.02, 0.01, 0.005, .0.004, 0.002, 0.0016,\right. \\ \left.0.001, 0.0007, 0.0004, 0.0001, 0.000025, 0.00001 \right\},
\end{multline}
as the lines appear from right to left. Again, the convergence is slow, especially around $\mu R=0$. We employ third-order polynomial fits in terms of $\mb$ at fixed $\mu R$ in the various regions of overlap of the curves. The extrapolation to $\mb =0$ results in the black curve that acts as an envelop of all the colored curves and represents the $a$-function in the continuum. The small gray part of the curve near the vertical axis is an extrapolation between the part computed numerically and $\mu R=0$, for which $a(0)=1$ is also obtained numerically in the massless theory. 

The result of our computation for the $a$-function in $3+1$ dimensions is a monotonic curve that interpolates between 1 and 0, similarly to the $c$-function in $1+1$ dimensions. The vanishing of the $a$-function  for $\mu R\to \infty$ confirms with high precision the presence of a term $\frac{1}{90}\ln \mu R$ in the entropy in this limit. A more dominant term $\sim \mu^2 R^2$ is also present, but it does not contribute to the $a$-function.

\section{Conclusions} \label{sec:concl}
 
The purpose of this work was to confirm the direct connection between entanglement entropy and the notion of irreversibility associated with the RG flow. We considered the simple example of a free massive scalar theory, which has a UV fixed point corresponding to the massless theory, while the field decouples in the IR, with no degrees of freedom surviving in this limit. According to \cite{Casini:2012ei,Casini:2017vbe}, the RG flow from the UV to the IR corresponds to the change of the entanglement  entropy for a spherical entangling surface as the radius grows from zero to infinity.  The $c$-function in $1+1$ dimensions and the $a$-function in $3+1$ dimensions are given by expressions \eqref{eq:dcr} and \eqref{eq:dar}, respectively \cite{Casini:2012ei,Casini:2017vbe,Abate:2024nyh}. Through an explicit calculation, starting from first principles and using analytical and numerical means, we confirmed this connection, both in $1+1$ and in $3+1$ dimensions. The application of  equations  \eqref{eq:dcr} and \eqref{eq:dar} to the entanglement entropy computed with the method of correlation functions has led to the results presented in figs. \ref{fig:c2} and \ref{fig:a}. Both functions are monotonic and vary continuously between 1 and 0, as expected. 
 
The calculation has revealed explicitly the difficulties in the calculation of $c$- and $a$-functions from first principles even in the simplest models. Analytical results, such as equations \eqref{eq:correctionzero} and  \eqref{eq:correctioninfinity}, are scarce and hard to obtain, existing mainly in lower dimensions. On the hand, the numerical approach is possible at the moment only for Gaussian states, limiting its application to non-interacting theories.
 
Despite these misgivings, very interesting calculations are feasible for free theories in gravitational backgrounds, especially those that exhibit horizons. In the case of de Sitter space a proposal for the extraction of $c$-, $F$- and $a$-functions from the entanglement entropy exists \cite{Abate:2024nyh}, while a calculation is possible through the extension of the analysis of \cite{Boutivas:2024sat,Boutivas:2024lts} to the massive field case.  The study of interacting theories, that exhibit RG flows between non-trivial fixed points, is a much harder task that requires the development of the formalism for studying entanglement beyond the case of Gaussian states.

\appendix
\section{Analytical Results}\label{sec:analytic}
In this appendix we derive the leading small $\mu R$ correction to the entanglement entropy of the massless theory. The calculation is performed using the methodology of \cite{Katsinis:2024gef}.

In order to probe the theory in the infinite size and massless limit, one has two choices. Either send the mass to zero with fixed system size $L$ and then the system size to infinity, or send the system size to infinity and then the mass to zero. Essentially, we have to chose whether to perform an expansion around $\mu L =0$ or $\mu L=\infty$.

Let us first analyze the $\mu L\ll 1$ regime. Expanding equations~\eqref{eq:kernel_OM_sin} and~\eqref{eq:kernel_OM_inv_sin} around $\mu=0$ yields
\begin{align}
	\Omega\left(x,x^\prime\right)&=\omega_{1}\left(x,x^\prime;L\right)+\delta\, \omega_{-1}\left(x,x^\prime;L\right)+\cdots\label{eq:ker_Om_massless},\\
	\Omega^{-1}\left(w,w^\prime\right)&=\omega_{-1}\left(x,w^\prime;L\right)+\delta\, \omega_{-3}\left(x,x^\prime;L\right)+\cdots,\label{eq:ker_Om_Inv_massless}
\end{align}
where the parameter $\delta$ of the expansion is
\begin{equation}\label{eq:delta_def}
\delta=-\frac{\mu^2L^2}{\pi^2}
\end{equation}
and $\omega_{-n}\left(x,x^\prime;L\right)$ is defined as
\begin{equation}
	\begin{split}\label{eq:omega_def}
		\omega_{-n} \left(x,x^\prime;L\right)&=\frac{2}{L}\sum_{k=1}^\infty\frac{1}{k^n}\sin\frac{k \pi x}{L}\sin\frac{k \pi x^\prime}{L}\\
		&=\frac{1}{2L} \left[\text{Li}_n\left(e^{\frac{i \pi  (x-x^\prime)}{L}}\right)-\text{Li}_n\left(e^{\frac{i \pi  (x+x^\prime)}{L}}\right)+\mathrm{c.c.}\right].
	\end{split}
\end{equation}
These are actually the kernels of the massless, finite-size theory. They are naturally expressed in terms of polylogarithms. This limit was studied extensively in \cite{Katsinis:2024gef}. In the infinite-size limit, the kernels $\omega_{1}$, $\omega_{-1}$, and $\omega_{-3}$ assume the form
\begin{align}
	\tilde{\omega}_{1}(x,x^\prime)&=\frac{L}{\pi^2}\left[\frac{1}{\left(x+x^\prime\right)^2}-\frac{1}{\left(x-x^\prime\right)^2}\right],\label{eq:om_p1_L_inf}\\
	\tilde{\omega}_{-1}(x,x^\prime)&=\frac{1}{L}\ln\frac{x+x^\prime}{\left\vert x-x^\prime\right\vert}\label{eq:om_m1_L_inf},\\
		\tilde{\omega}_{-3}(x,x^\prime;\tilde{L})&=\frac{\pi^2}{2L^3}\left[\left(x-x^\prime\right)^2\ln\frac{\left\vert x-x^\prime\right\vert}{\tilde{L}}-\left(x+x^\prime\right)^2\ln\frac{(x+x^\prime)}{\tilde{L}}\right],\label{eq:om_m3_L_inf}
\end{align}
where $\tilde{\omega}$ denotes the infinite-size kernels and $\tilde{L}=e^{3/2}L/\pi$.

One can use perturbation theory around the massless limit, along the lines of \cite{Boutivas:2025rdf}\footnote{Reference \cite{Boutivas:2025rdf} concerns the entanglement entropy of a scalar field in $\mathbb{R}\times$S$^3$. In order to make contact with the present work, one has to set $\theta_M=L/a$, $\theta_R=R/a$ and send $a$ to infinity. Then, the calculation is literally the same.}. To order $\delta$, the product of $\Omega^{-1}\left(x,y\right)$ and $\Omega\left(y,x^\prime\right)$ reads
\begin{multline}\label{eq:product_Omegas}
	\Omega^{-1}\left(x,y\right)\Omega\left(y,x^\prime\right)= \omega_{-1} \left(x,y\right)\omega_{1} \left(y,x^\prime\right)\\
	+\frac{1}{2}\delta\left[\tilde{\omega}_{-3} \left(x,y;\tilde{L}\right)\tilde{\omega}_{1} \left(y,x^\prime\right)-\tilde{\omega}_{-1} \left(x,y\right)\tilde{\omega}_{-1} \left(y,x^\prime\right)\right]+\mathcal{O}\left(\delta^2\right).
\end{multline}	
Writing $S=S^{(0)}+\delta S^{(1)}+\cdots$ it turns out that  \cite{Boutivas:2025rdf}
\begin{equation}\label{eq:S1EE}
\delta\, S^{(1)}=-\frac{1}{6}\mu^2R^2\left(\ln \frac{L}{2 \pi R}+\frac{4}{3}\right)=-\frac{1}{6}\mu^2R^2\left(\ln \frac{\tilde{L}}{2R}-\frac{1}{6}\right).
\end{equation}
The total entanglement entropy is given by
\begin{equation} \label{eq:finalresult}
	S=\frac{1}{6}\ln\frac{2R}{\epsilon}-\frac{1}{6}R^2\mu^2\left(\ln \frac{L}{2 \pi R}+\frac{4}{3}\right)+\mathcal{O}\left(\delta^2\right)+\mathcal{O}\left(\delta \frac{R^3}{L^3}\right),
\end{equation}
where the leading term is the well-known result of \cite{Holzhey:1994we,Calabrese:2004eu}. We stress that this expression is valid for $\mu L \ll 1$ and $R\ll L$.

It can be shown that equations \eqref{eq:kernel_OM_sin} and \eqref{eq:kernel_OM_inv_sin} can also be written in the form
\begin{align}
\Omega(x,x^\prime)&=-\frac{\mu}{\pi}\sum_{p=-\infty}^\infty\left(\frac{K_{1}\left(\mu\vert x_-\vert\right)}{\vert x_-\vert}-\frac{K_{1}\left(\mu\vert x_+\vert\right)}{\vert x_+\vert}\right)\label{eq:Omega_pos_con_1},\\
\Omega^{-1}(x,x^\prime)&=\frac{1}{\pi}\sum_{p=-\infty}^\infty\bigg(K_{0}\left(\mu\vert x_-\vert\right)-K_{0}\left(\mu\vert x_+\vert\right)\bigg)\label{eq:Omega_neg_con_1},
\end{align}
where $x_\pm=x \pm x^\prime-2pL$. Details are provided in appendix B.2 of \cite{Katsinis:2024gef}. This form is more appropriate for the other interesting limit, i.e., $\mu L\gg 1$. It turns out that in this limit only the $p=0$ terms contribute to $\Omega(x,x^\prime)$ and $\Omega^{-1}(x,x^\prime)$ and we have\footnote{The same result can be obtained by considering the $L\rightarrow\infty$ limit of \eqref{eq:kernel_OM_sin} and \eqref{eq:kernel_OM_inv_sin}. In this limit $k\pi/L$ becomes a continuous variable and the above equations assume the form of a cosine Fourier transform that gives exactly the same result.}
\begin{align}
\Omega(x,x^\prime)&=\frac{\mu}{\pi}\left(\frac{K_{1}\left(\mu\vert x+x^\prime\vert\right)}{\vert x+x^\prime\vert}-\frac{K_{1}\left(\mu\vert x-x^\prime\vert\right)}{\vert x-x^\prime\vert}\right)\label{eq:Omega_pos_con_3},\\
\Omega^{-1}(x,x^\prime)&=\frac{1}{\pi}\left[K_{0}\left(\mu\vert x-x^\prime\vert\right)-K_{0}\left(\mu\vert x+x^\prime\vert\right)\right]\label{eq:Omega_neg_con_3}.
\end{align}

Next, we consider the limit $\mu \rightarrow 0$ using
\begin{align}
K_{0}(x)&=-\left(1+\frac{x^2}{4}+\mathcal{O}\left(x^4\right)\right) \left(\ln\frac{x}{2}+\gamma \right)+\frac{x^2}{4}+\mathcal{O}\left(x^4\right),\label{eq:K0_zero}\\
K_{1}(x)&=\left(\frac{x}{2}+\mathcal{O}\left(x^4\right)\right) \left(\ln\frac{x}{2}+\gamma \right)+\frac{1}{x}-\frac{x}{4}+\mathcal{O}\left(x^3\right),\label{eq:K1_zero}
\end{align}
where $\gamma$ is the Euler–Mascheroni constant. We obtain
\begin{align}
\Omega(x,x^\prime)&=\frac{1}{\pi}\left[\frac{1}{(x+x^\prime)^2}-\frac{1}{(x-x^\prime)^2}\right]+\frac{\mu ^2 }{2\pi}\ln\left\vert\frac{x+x^\prime}{x-x^\prime}\right\vert+\cdots\label{eq:Omega_pos_Expansion},\\
\Omega^{-1}(x,x^\prime)&=\frac{1}{\pi}\ln\left\vert\frac{x+x^\prime}{x-x^\prime}\right\vert-\frac{\mu ^2}{4 \pi}\left[(x-x^\prime)^2 \ln\frac{\tilde{\mu}\vert x-x^\prime\vert}{2}-\left(x^\prime\rightarrow-x^\prime\right)\right]+\cdots\label{eq:Omega_neg_Expansion},
\end{align}
where $\tilde{\mu}=\mu e^{\gamma -1}$. Notice that after sending the system size to infinity, we have implicitly introduced the IR cutoff $1/\mu$, which is required for the validity of the expansions \eqref{eq:K0_zero} and \eqref{eq:K1_zero}\footnote{In this work, we have defined the overall system as the half-line $[0,\infty)$. Had we considered the whole line, the kernels would assume the form appearing in \cite{Callan:1994py}, where the same approach is used in order to probe the massless limit.}.

Using the notation of equations \eqref{eq:om_p1_L_inf}, \eqref{eq:om_m1_L_inf} and \eqref{eq:om_m3_L_inf}, we have
\begin{multline}
\Omega^{-1}\left(x,y\right)\Omega\left(y,x^\prime\right)= \omega_{-1} \left(x,y\right)\omega_{1} \left(y,x^\prime\right)\\+\frac{\delta}{2}\left[\tilde{\omega}_{-3}(x,y;2/\tilde{\mu})\tilde{\omega}_{1}(y,x^\prime)-\tilde{\omega}_{-1}(x,y)\tilde{\omega}_{-1}(y,x^\prime)\right],
\end{multline}
where $\delta$ is defined in equation \eqref{eq:delta_def}. Comparing with equation \eqref{eq:product_Omegas} we conclude that the entanglement entropy is given by equation \eqref{eq:S1EE} with $\tilde{L}\rightarrow 2/\tilde{\mu}$, namely
\begin{equation}
	\delta\, S^{(1)}(\mu R)=\frac{1}{6}\mu^2R^2\left(\ln\left(\tilde{\mu}R\right)+\frac{1}{6}\right)=\frac{1}{6}\mu^2R^2\left(\ln\left(\mu R\right)+\gamma -\frac{5}{6}\right).
\end{equation}
In analogy with equation \eqref{eq:finalresult}, we have
\begin{equation} 
	S=\frac{1}{6}\ln\frac{2R}{\epsilon}+\frac{1}{6}\mu^2R^2\left(\ln\left(\mu R\right)+\gamma -\frac{5}{6}\right)+\mathcal{O}\left(\delta^2\right)+\mathcal{O}\left(\delta \frac{R^3}{L^3}\right).
\end{equation} 

\bibliographystyle{JHEP} % or JHEP.bst if available
\bibliography{afunction_bib}

\end{document}